\documentclass[a4paper]{jpconf}
\usepackage[dvipdfm]{graphicx}
\usepackage{bm}
\usepackage{cite}
\usepackage{xcolor}

\begin{document}
\title{Confined states in the tight-binding model on the hexagonal golden-mean tiling}

\author{Toranosuke Matsubara${}^1$, Akihisa Koga${}^1$ and Sam Coates${}^2$}

\address{${}^1$Department of Physics, Tokyo Institute of Technology, Meguro, Tokyo 152-8551, Japan\\ ${}^2$Department of Materials Science and Technology, Tokyo University of Science, Katsushika, Tokyo 125-8585, Japan}

\ead{matsubara@stat.phys.titech.ac.jp}


\begin{abstract}
  We study the tight-binding model with two distinct hoppings $(t_L, t_S)$
  on the two-dimensional hexagonal golden-mean tiling
  and examine the confined states with $E=0$,
  where $E$ is the eigenenergy.  
  Some confined states found in the case $t_L=t_S$
  are exact eigenstates even for the system with $t_L\neq t_S$,
  where their amplitudes are smoothly changed.
  By contrast, the other states are no longer eigenstates of
  the system with $t_L\neq t_S$.
  This may imply the existence of macroscopically degenerate states which are characteristic of the system with $t_L=t_S$, and that a discontinuity appears in
  the number of the confined states in the thermodynamic limit.
\end{abstract}

\section{Introduction}
Quasicrystals have attracted much interest
since the discovery of the quasicrystalline phase of
Al-Mn alloy~\cite{Shechtman_1984}.
Among them, electron correlations in quasicrystals
have actively been discussed after the observation of quantum critical behavior
in Au-Al-Yb~\cite{Deguchi}.
Recently, long-range correlations have been observed --
such as superconductivity in the Al-Zn-Mg quasicrystal~\cite{Kamiya_2018} 
and the ferromagnetically ordered states in the Au-Ga-Gd and Au-Ga-Tb
quasicrystals~\cite{Tamura_2021}.
These experiments have necessarily stimulated further theoretical investigations
on electron correlations in quasiperiodic tilings~\cite{Watanabe_2013,Takemori_2015,Takemura_2015,Andrade_2015,Otsuki_2016,Sakai_2017,Ara_2019,Varjas_2019,Duncan_2020,Ghadimi_2021}.

A simple example of such investigations is the study of magnetically ordered states
in the Hubbard model on the bipartite quasiperiodic tilings, i.e., the Penrose~\cite{Jag_2007,Koga_2017},
Ammann-Beenker~\cite{Jag_1997,Wessel_2003,Koga_2020},
and Socolar dodecagonal tilings~\cite{Koga_2021}.
Non--interacting systems on the above tilings have
a common feature in the density of states, namely,
macroscopically degenerate states at $E=0$,
so-called confined states.
These play an essential role
for stabilizing the magnetically ordered states, in particular,
in the weak coupling regime.
Therefore, to understand magnetic properties of quasiperiodic tilings,
it is important to examine
their confined states under the tight-binding
model~\cite{Kohmoto,Arai_1988,Koga_2017,Koga_2020,Oktel,Koga_2021}.

\begin{figure}[htb]
  \begin{center}
    \includegraphics[width=\linewidth]{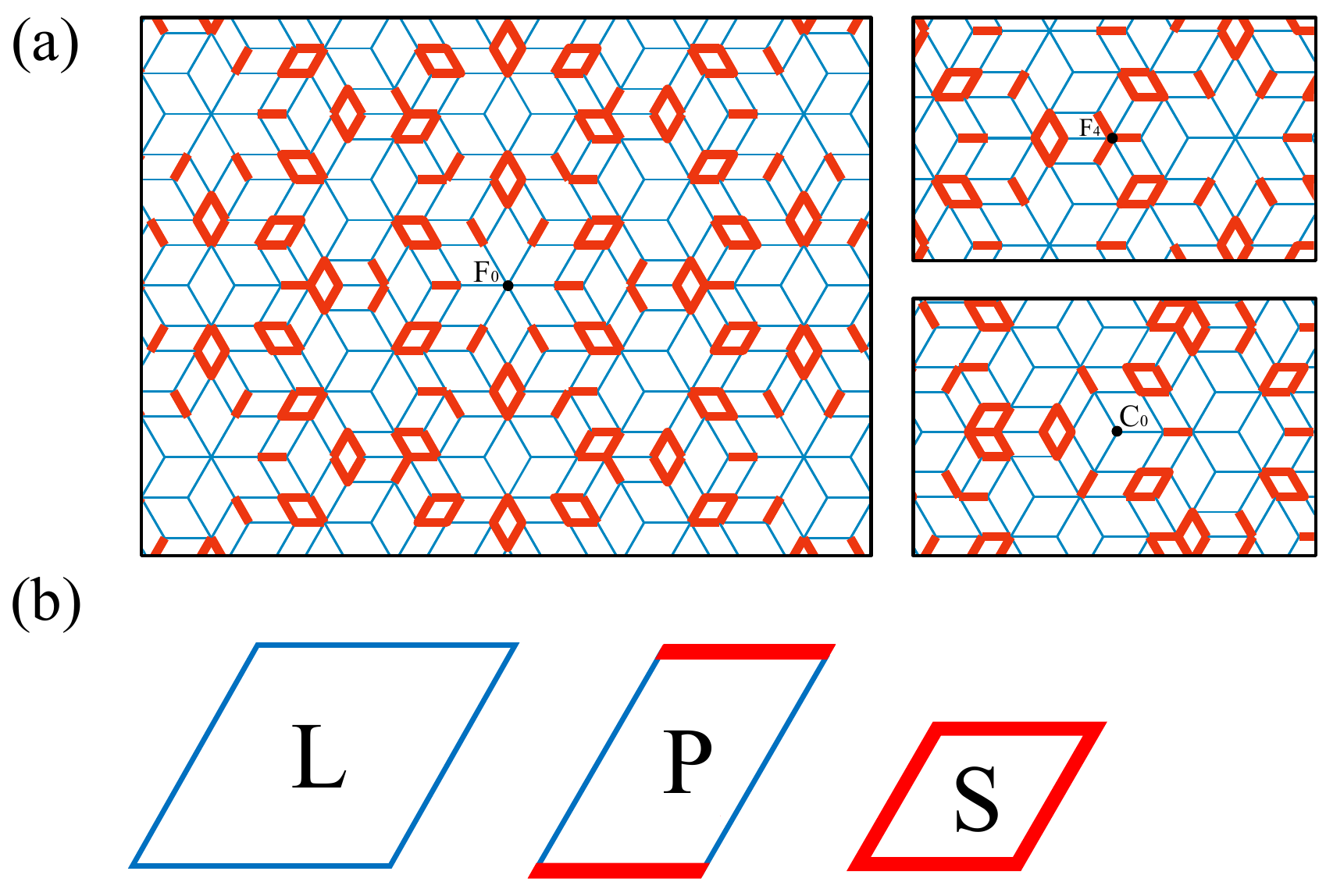}
    \caption{(a) Hexagonal golden-mean tiling\cite{Coates_2022}.
      Blue thin and red bold lines indicate
      the hopping integrals $t_L$ and $t_S$ defined on the
      long and short length edges.
      Solid circles indicate the F$_0$, F$_4$, and C$_0$ vertices.
      (b) Large rhombus, parallelogram, and small rhombus.
    }
    \label{tile}
  \end{center}
\end{figure}

Recently, the hexagonal golden-mean tiling
has been introduced~\cite{Coates_2022}, with a section of the tiling and its constituent tiles shown in Fig.~\ref{tile}.
This tiling is composed of large rhombuses, parallelograms,
and small rhombuses, and, one of its important features is the existence of
two length scales
which is in contrast to the Penrose, Ammann-Beenker,
and Socolar dodecagonal tilings.
In our previous paper~\cite{Koga_2022}, we have considered the vertex model
on the hexagonal golden-mean tiling,
where the hopping integral on each edge is assumed to be equivalent.
However, it is also instructive to clarify confined state properties
in the tight-binding model with distinct hoppings.

The paper is organized as follows: in Sec.~\ref{sec:TB},
we introduce the tight-binding model on the hexagonal golden-mean tiling
and show the density of states.
In Sec.~\ref{sec:conf},
we discuss confined state properties in the model and compare our results to our previous work and the Penrose tiling.
A summary is given in the last section.

\section{Tight-binding model on the hexagonal golden-mean tiling}\label{sec:TB}

Here we briefly summarise relevant properties of the hexagonal golden-mean tiling as introduced in the original paper~\cite{Coates_2022}. The tiling can be generated using deflation rules for eight distinct directed tiles, and there are 32 allowed vertex configurations.
Important for further discussion are the F$_0$, F$_4$, and C$_0$ vertices. The F$_0$ vertex is located at the {\color{black} centre} of
six adjacent large rhombuses and locally has 
6-fold rotational symmetry, while the F$_4$, and C$_0$ vertices
locally have 3-fold rotational symmetry.

As the tiling is multi-length-scale, we can introduce two kinds of hopping integrals in the tight-binding model.
The tight-binding model with two distinct hoppings is given as
\begin{eqnarray}
H =  -t_L\sum_{(ij)} ( c^\dagger_{i} c_{j} + {\rm H.c.}  )-t_S\sum_{\langle ij\rangle} ( c^\dagger_{i} c_{j} + {\rm H.c.}  )  ,
\label{TBmodel}
\end{eqnarray}
where $c_{i}\,(c^\dagger_{i})$ annihilates (creates) an electron at the
$i$th site.
$t_L(t_S)$ denotes the transfer integral between the nearest neighbour pairs
on the long (short) bonds,
which are shown as the thin (bold) lines in Fig.~\ref{tile}(a).
Then, we examine the density of states as
\begin{eqnarray}
  D(E)=\frac{1}{N}\sum_i\delta\left(E-\epsilon_i\right),
\end{eqnarray}
where $\epsilon_i$ is the $i$th eigenvalue of the Hamiltonian eq.~(\ref{TBmodel})
and $N$ is the number of sites.
\begin{figure}[htb]
  \begin{center}
    \includegraphics[width=\linewidth]{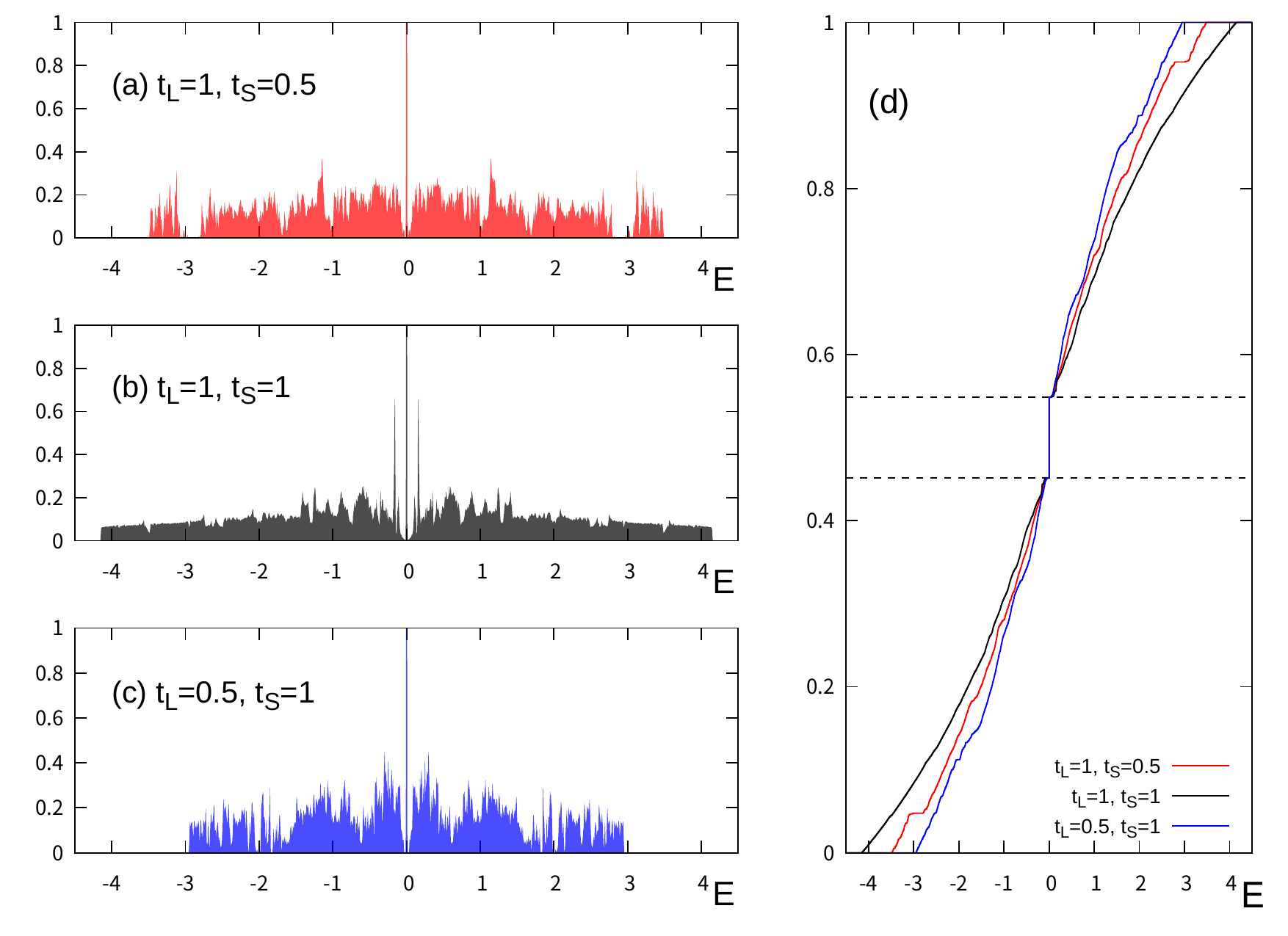}
    \caption{Density of states for the tight-binding model on the hexagonal
      golden-mean tiling when $(t_L, t_S)=$ (a)\; (1,0.5), (b)\; (1,1),
      and (c)\; (0.5,1).
      (d) shows the integrated densities of states $I(E)=\int^E_{-\infty} D(\epsilon)d\epsilon$.
    }
    \label{dos}
  \end{center}
\end{figure}
Figures~\ref{dos}(a), (b), and (c) show the density of states for the tight-binding models
with $(t_L, t_S)=(1,0.5), (1,1)$, and $(0.5,1)$,
by numerically diagonalizing the Hamiltonian with $N=448,213$ under the open boundary condition.
{\color{black}
Since the system is bipartite, the density of states is symmetric along $E=0$.
}
We clearly find a delta-function-like peak at $E=0$ for each case, which implies the existence of confined states in the model.
We also demonstrate that the number of the eigenstates with $E=0$ is not changed as $t_L(t_S)$ is changed,
as shown by the integrated density of states in Fig.~\ref{dos}(d).
However, it is not clear that
each eigenstate with $E=0$ is always confined,
since certain confined states may be eigenstates
only under special conditions.
To clarify this, we now carefully consider the confined states in the system and compare them to the states found under $t_L=t_S$.

\section{Confined state properties}\label{sec:conf}
The macroscopically degenerate states at $E=0$ can be described
in a simple form by considering their appropriate linear combination.
Some confined states for the case $t_L=t_S$
have explicitly been shown in the previous paper~\cite{Koga_2022},
where each confined state has amplitudes in the finite region {\color{black} of real space}.
Since a certain finite region is repeated quite regularly in the tiling, the number of confined states is infinite,
leading to the delta-function-like peak in the density of states.

\begin{figure}[htb]
  \begin{center}
    \includegraphics[width=\linewidth]{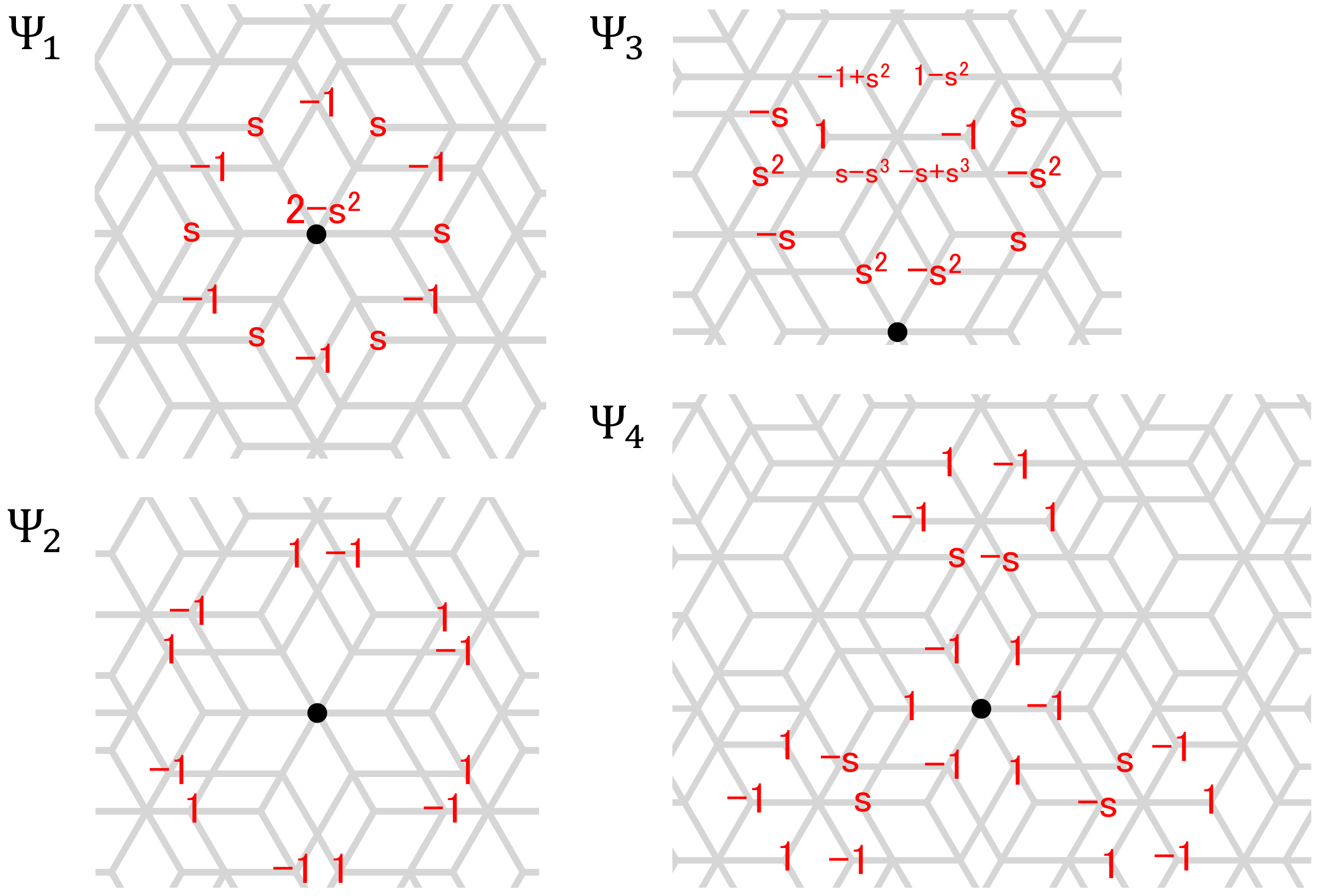}
    \caption{Four confined states around the F$_0$ vertices, {\color{black} shown as solid circles}.
      The values at the vertices represent the amplitudes of
      the confined states with $s=t_S/t_L$.
    }
    \label{conf states conserve}
  \end{center}
\end{figure}

We now consider the tight-binding model eq. (\ref{TBmodel}) with $t_L\neq t_S$.
Some simple examples around $F_0$ vertices
are shown in Fig.~\ref{conf states conserve},
where the $F_0$ vertex is located
at the {\color{black} centre} of six adjacent large rhombuses. 
Two confined states $\Psi_1$ and $\Psi_2$ are located
inside almost the same region.
Since both states are reduced versions of the simple ones for $t_L=t_S$
obtained in the previous paper~\cite{Koga_2022},
we can say that both confined states are not changed
in their properties, and each fraction is thereby given as $1/(4\tau^8)$.
Namely, these confined states have amplitudes
in the distinct sublattices.
This means that the staggered magnetization
is induced by the infinitesimal Coulomb interactions
if one considers the Hubbard model.
Similar behavior appears in the confined states $\Psi_3$ and $\Psi_4$.
However, each of the previously obtained confined states with $t_L=t_S$ 
are not necessarily shared in the model with $t_L\neq t_S$.
\begin{figure}[htb]
  \begin{center}
    \includegraphics[width=\linewidth]{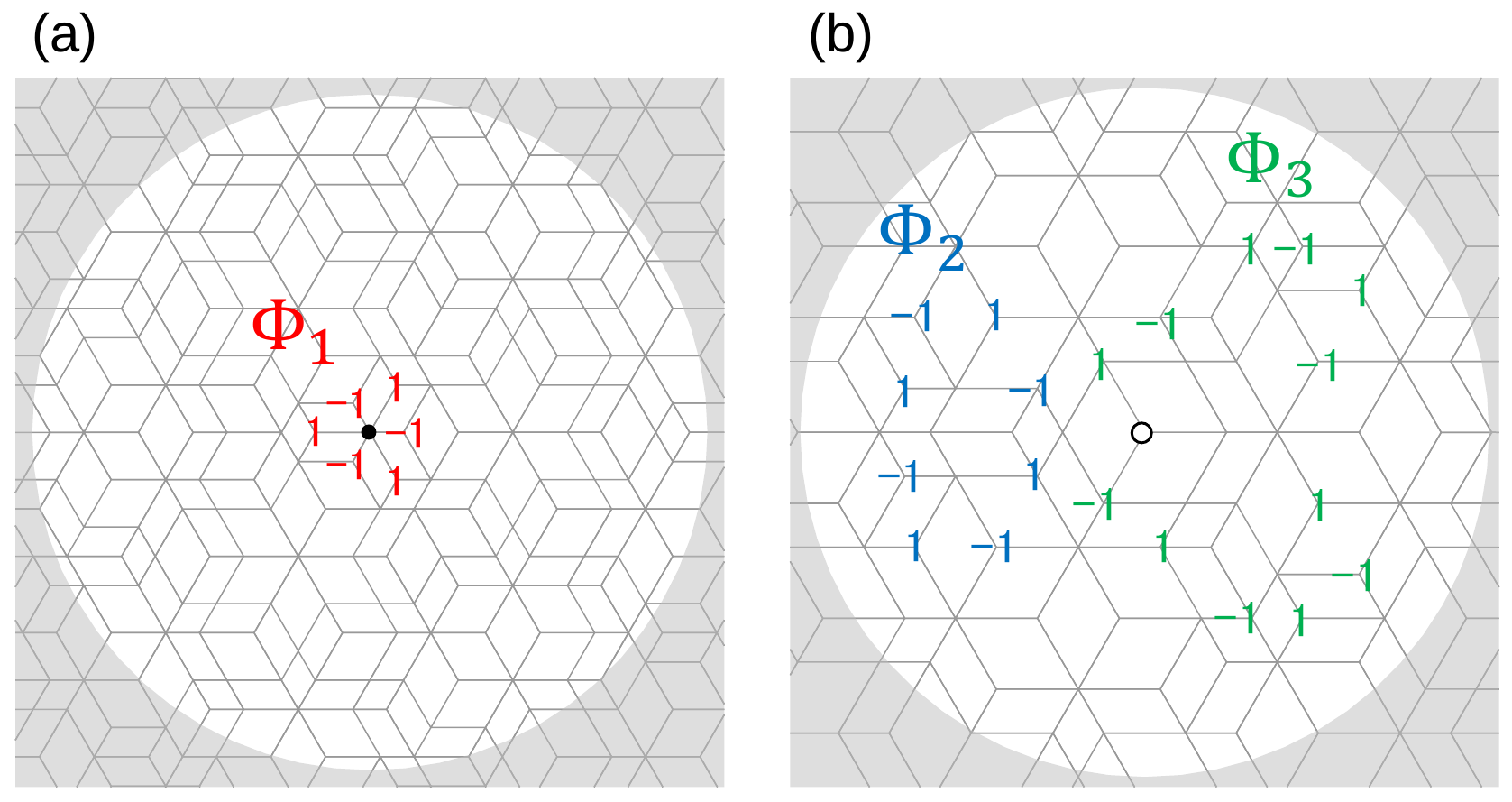}
    \caption{(a) Confined state $\Phi_1$ around the F$_4$ vertex, {\color{black} shown as a solid circle}
      and (b) two confined states $\Phi_2$ and $\Phi_3$ around the C$_0$ vertex, {\color{black} shown as an open circle},
      for the tight-binding model with $t_L=t_S$~\cite{Koga_2022}.
      The numbers at the vertices colored by the red, blue, and green
      represent the amplitudes of the confined states $\Phi_1$, $\Phi_2$, $\Phi_3$,
      respectively.
    }
    \label{break}
  \end{center}
\end{figure}
Figure~\ref{break}(a) shows the confined state around the F$_4$ vertex
for the model with $t_L=t_S$,
which was labelled as $\Phi_1$ in our paper~\cite{Koga_2022}.
When $t_L\neq t_S$, this state is not confined around the F$_4$ vertex; in fact, we can find no eigenstates with $E=0$
in the circular region shown in Fig.~\ref{break}(a)
by numerically diagonalizing the Hamiltonian,
\begin{eqnarray}
H'=H+\sum_iV_ic_i^\dag c_i,
\end{eqnarray}
where $V_i$ is the potential in the outside of the circular region.
We also examine the confined states around the C$_0$ vertex.
In the case of $t_L=t_S$, there are five confined states
(three $\Phi_2$ and two $\Phi_3$)
in the circular region shown in Fig.~\ref{break}(b).
When the ratio is away from the condition $t_L = t_S$,
the number of the confined states is changed from five to three.
However, the three confined states are difficult to describe by the simple form (linear combination).
We can therefore say that two confined states exist
which are inherent in the special case with $t_L=t_S$,
in contrast to the altered confined states discussed above.
This is not inconsistent with the fact that, in the large cluster treatment,
the eigenstates with $E=0$ are not changed by varying $t_L/t_S$
{\color{black} since some eigenstates with $E=0$ exist with amplitudes at the edges of the system}.
Therefore, magnetic properties in the weak coupling limit
should depend on the ratio $t_L/t_S$.

\begin{figure}[htb]
  \begin{center}
    \includegraphics[width=\linewidth]{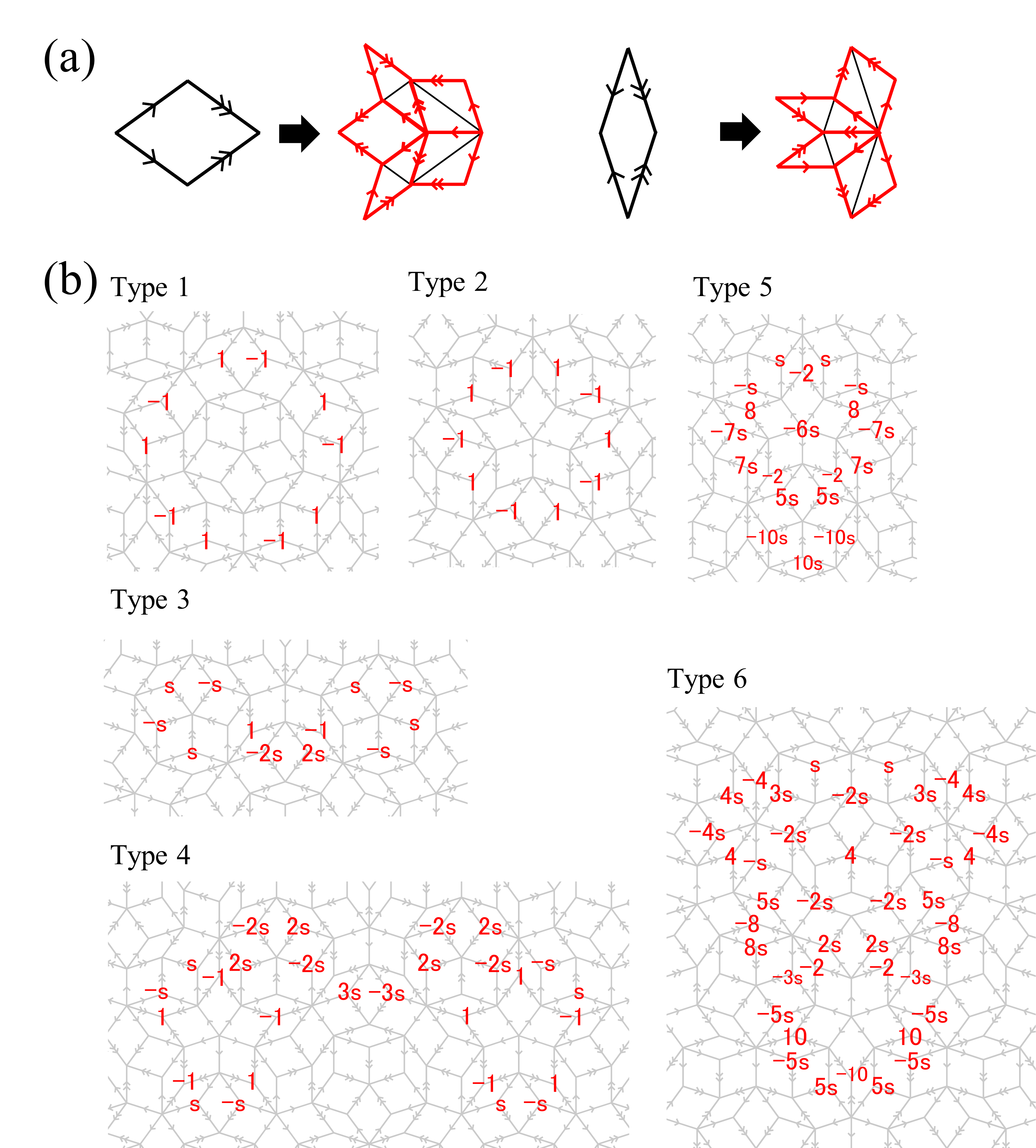}
    \caption{(a) Deflation rule of fat and skinny rhombuses in the Penrose tiling.
      (b) Six types of the confined states in the tight-binding model
      on the vertices of the Penrose tiling.
      The numbers at the vertices represent
      the amplitudes of confined states with {\color{black} $s=t_2/t_1$}.}
    \label{penrose and two TI model}
  \end{center}
\end{figure}

We also compare our results to the confined state properties in the Penrose tiling.
The Penrose tiling is composed of fat and skinny rhombuses
with a unit length scale.
However, we know from edge-matching rules that there are two kinds of edges in the deflation rule
for the directed fat and skinny rhombuses,
as shown in Fig.~\ref{penrose and two TI model}.
This allows us to introduce distinct hoppings $(t_1,t_2)$
in the tight-binding model on the Penrose tiling,
where $t_1 (t_2)$ is the hopping integral
defined on the bond with a single (double) arrow. 
The vertex model on the Penrose tiling has been examined in detail and  
there are six types of confined states
when $t_1=t_2$~\cite{Kohmoto,Arai_1988}.
We confirm that six states 
are still confined in the same region, with some modifications in the amplitudes,
as shown in Fig.~\ref{penrose and two TI model}.
This means that the fraction of the confined states is not changed
by the introduction of the distinct hopping integrals
in the tight-binding model on the Penrose tiling.
This is in contrast to the case in the hexagonal golden-mean tiling as
discussed above.
{\color{black}
We also note that the two hopping integrals cannot be introduced 
into the Ammann-Beenker and Socolar dodecagonal tilings
since the edges of their tiles are not divided into two groups, due to their geometry.
Therefore, we can say that the confined state properties in the hexagonal golden-mean tiling
are distinct from those in the others.
}


\section{Summary}
We have investigated the tight-binding model with two distinct hoppings
on the two-dimensional hexagonal golden-mean tiling.
Examining the confined states,
we have shown that
some confined states found in the case $t_L=t_S$
are also exact eigenstates for the system, even when $t_L\neq t_S$.
Similarly, some states are no longer eigenstates of
the system when $t_L\neq t_S$.
This may imply the existence of macroscopically degenerate states which are
characteristic of the system with $t_L=t_S$.
These properties are distinct from those in the tight-binding models
on the conventional quasiperiodic tilings such as the Penrose, Ammann-Beenker,
and Socolar dodecagonal tilings.
{\color{black}
It is an important problem to clarify magnetic properties
in the Hubbard model since the confined states with $E=0$ play an essential role
for the weak coupling limit, which will be discussed in the future.
}

\section*{Acknowledgements}
This work was supported by Grant-in-Aid for Scientific Research from JSPS, KAKENHI Grants
 No. JP17K05536, JP19H05821, JP21H01025, JP22K03525 (A.K.), and No. JP19H05817 and No. JP19H05818 (S.C.). 

\section*{References}
\bibliographystyle{iopart-num}
\bibliography{refs}

\providecommand{\newblock}{}
\begin{thebibliography}{10}
\expandafter\ifx\csname url\endcsname\relax
  \def\url#1{{\tt #1}}\fi
\expandafter\ifx\csname urlprefix\endcsname\relax\def\urlprefix{URL }\fi
\providecommand{\eprint}[2][]{\url{#2}}

\bibitem{Shechtman_1984}
Shechtman D, Blech I, Gratias D and Cahn J~W 1984 {\em Physical Review
  Letters\/} {\bf 53} 1951

\bibitem{Deguchi}
Deguchi K, Matsukawa S, Sato N~K, Hattori T, Ishida K, Takakura H and Ishimasa
  T 2012 {\em Nature Materials\/} {\bf 11} 1013^^e2^^80^^931016 ISSN 1476-1122

\bibitem{Kamiya_2018}
Kamiya K, Takeuchi T, Kabeya N, Wada N, Ishimasa T, Ochiai A, Deguchi K, Imura
  K and Sato N 2018 {\em Nature Communications\/} {\bf 9} 1--8

\bibitem{Tamura_2021}
Tamura R, Ishikawa A, Suzuki S, Kotajima T, Tanaka Y, Seki T, Shibata N, Yamada
  T, Fujii T, Wang C~W {\em et~al.\/} 2021 {\em Journal of the American
  Chemical Society\/} {\bf 143} 19938--19944

\bibitem{Watanabe_2013}
Watanabe S and Miyake K 2013 {\em Journal of the Physical Society of Japan\/}
  {\bf 82} 083704 ISSN 0031-9015

\bibitem{Takemori_2015}
Takemori N and Koga A 2015 {\em Journal of the Physical Society of Japan\/}
  {\bf 84} 023701 ISSN 0031-9015

\bibitem{Takemura_2015}
Takemura S, Takemori N and Koga A 2015 {\em Physical Review B\/} {\bf 91}
  165114

\bibitem{Andrade_2015}
Andrade E~C, Jagannathan A, Miranda E, Vojta M and Dobrosavljevi{\'c} V 2015
  {\em Physical Review Letters\/} {\bf 115} 036403

\bibitem{Otsuki_2016}
Otsuki J and Kusunose H 2016 {\em Journal of the Physical Society of Japan\/}
  {\bf 85} 073712 ISSN 0031-9015

\bibitem{Sakai_2017}
Sakai S, Takemori N, Koga A and Arita R 2017 {\em Physical Review B\/} {\bf 95}
  024509

\bibitem{Ara_2019}
Ara\'ujo R~N and Andrade E~C 2019 {\em Physical Review B\/} {\bf 100} 014510

\bibitem{Varjas_2019}
Varjas D, Lau A, P\"oyh\"onen K, Akhmerov A~R, Pikulin D~I and Fulga I~C 2019
  {\em Physical Review Letters\/} {\bf 123} 196401

\bibitem{Duncan_2020}
Duncan C~W, Manna S and Nielsen A~E~B 2020 {\em Physical Review B\/} {\bf 101}
  115413

\bibitem{Ghadimi_2021}
Ghadimi R, Sugimoto T, Tanaka K and Tohyama T 2021 {\em Physical Review B\/}
  {\bf 104} 144511

\bibitem{Jag_2007}
Jagannathan A, Szallas A, Wessel S and Duneau M 2007 {\em Physical Review B\/}
  {\bf 75} 212407

\bibitem{Koga_2017}
Koga A and Tsunetsugu H 2017 {\em Physical Review B\/} {\bf 96} 214402

\bibitem{Jag_1997}
Jagannathan A and Schulz H~J 1997 {\em Physical Review B\/} {\bf 55}
  8045^^e2^^80^^938048

\bibitem{Wessel_2003}
Wessel S, Jagannathan A and Haas S 2003 {\em Physical Review Letters\/} {\bf
  90} 177205

\bibitem{Koga_2020}
Koga A 2020 {\em Physical Review B\/} {\bf 102} 115125

\bibitem{Koga_2021}
Koga A 2021 {\em Materials Transactions\/} {\bf 62} 360--366

\bibitem{Kohmoto}
Kohmoto M and Sutherland B 1986 {\em Physical Review B\/} {\bf 34} 3849--3853

\bibitem{Arai_1988}
Arai M, Tokihiro T, Fujiwara T and Kohmoto M 1988 {\em Physical Review B\/}
  {\bf 38} 1621

\bibitem{Oktel}
Oktel M~O 2021 {\em Physical Review B\/} {\bf 104}(1) 014204

\bibitem{Coates_2022}
Coates S, Lifshitz R, Koga A, McGrath R, Sharma H~R and Tamura R 2022 {\em
  arXiv preprint arXiv:2201.11848\/}

\bibitem{Koga_2022}
Koga A and Coates S 2022 {\em Physical Review B\/} {\bf 105} 104410

\end{thebibliography}

\end{document}